# Enforced symmetry breaking for anomalous valley Hall effect in two-dimensional hexagonal lattices

Yongqian Zhu,[1,2] Jia-Tao Sun,[3,*] Jinbo Pan,[1,2,4] Jun Deng,[1] and Shixuan Du[1,2,4,*]

[1]*Beijing National Laboratory for Condensed Matter Physics, Institute of Physics, Chinese Academy of Sciences, Beijing 100190, China*
[2]*University of Chinese Academy of Sciences, Chinese Academy of Sciences, Beijing 100190, China*
[3]*School of Integrated Circuits and Electronics, MIIT Key Laboratory for Low-Dimensional Quantum Structure and Devices, Beijing Institute of Technology, Beijing 100081, China*
[4]*Songshan Lake Materials Laboratory, Dongguan 523808, China*

The anomalous valley Hall effect (AVHE) is a pivotal phenomenon that allows for the exploitation of the valley degree of freedom in materials. A general strategy for its realization and manipulation is crucial for valleytronics. Here, by considering all possible symmetries, we propose general rules for the realization and manipulation of AVHE in two-dimensional hexagonal lattices. The realization of AVHE requires breaking the enforced symmetry that is associated with different valleys or reverses the sign of Berry curvature. Further manipulation of AVHE requires asymmetry operators connecting two states with opposite signs of Berry curvature. These rules for realizing and manipulating AVHE are extendable to generic points in momentum space. Combined with first-principles calculations, we realize the controllable AVHE in four representative systems, i.e., monolayer $AgCrP_2Se_6$, $CrOBr$, $FeCl_2$ and bilayer $TcGeSe_3$. Our work provides symmetry rules for designing valleytronic materials that could facilitate the experimental detection and realistic applications.

*Introduction.*—Valleytronics is a branch of electronics that utilizes the valley degree of freedom (DOF) to process or store information [1-3]. The valley DOF specifies the valley occupied by the electron and possesses unique characteristics. Compared to charge, it does not produce direct Joule heating. Compared to spin, it can further suppress electron scattering because the intervalley scattering can be suppressed [4-6]. Coupling the valley DOF to Berry curvature can give rise to the anomalous valley Hall effect (AVHE), enabling the electrical readout of valley DOF. The realization and manipulation of AVHE are crucial for valleytronic applications [7,8]. Researchers have found that realizing the AVHE requires breaking certain symmetries, such as inversion, time-reversal and mirror symmetries, which enforce the valley degeneracy or zero net Berry curvature [4,5,7-24]. However, it lacks a general symmetry rule for achieving the AVHE, which hinders the development of a general strategy for designing AVHE systems.

The manipulation of AVHE requires reversing the sign of Berry curvature, which can be achieved through various means such as reversing the direction of magnetic moment [8,25] or external electric field [7], reversing the ferroelectric polarization [26] or changing the stacking order [19,27-30]. Among them, the mechanism of Berry curvature reversal by switching the stacking order remains unclear. Researchers found that two states of monolayer $VSe_2$ connected by time-reversal operator with opposite magnetic moments exhibit opposite valley energy splitting and opposite Berry curvature distributions [8], suggesting their valley related properties are interconnected through the time-reversal operator. The connection between configurations and valley properties inspires us to consider whether an asymmetric transformation between two configurations of a bilayer structure will lead to Berry curvature reversal. Moreover, previous studies on AVHE were usually limited to the high symmetric momentum points without considering the generic momentum points, which significantly narrows down the candidate materials for valleytronics.

In this Letter, based on systematic group theory analysis, we propose general symmetry rules for realizing AVHE and switching the valley DOF for 2D hexagonal systems. All the double magnetic space groups permitting AVHE are identified. We reveal that the valley DOF associating two switchable states via the asymmetric connection operation can be switched. Interestingly, the connection operation also determines the coupling between valley, spin and layer DOFs, as well as the magnetoelectric coupling. Moreover, we generalize the location of valley to generic momentum points. Combined with first-principles calculations, we predict the reversible manipulation of AVHE in four representative systems, i.e., monolayer $AgCrP_2Se_6$, $CrOBr$, $FeCl_2$ and bilayer $TcGeSe_3$.

*Group theory analysis.*—We consider a 2D hexagonal system with spin-orbit coupling, whose valleys are located at the high symmetric K point in reciprocal space. We denote the symmetry operation as $\hat{R}_m$, where $\hat{R}_m$ belongs to the double magnetic space group (DMSG) of the system and the subscript "m" of $\hat{R}_m$ denotes "magnetic" [31-33]. AVHE requires not only a valley polarization (i.e., valley splitting) but also a nonzero net Berry curvature associated with valleys [7,8,19]. As shown in the upper panel of Fig. 1, if existing a symmetry operator which connects two valleys at $+K$ and $-K$ points or reverses the sign of Berry curvature ($\Omega$), valley degeneracy or zero net Berry curvature will be enforced (Table I and Sec. 1 of the Supplemental Material [34]). We refer to this symmetry as enforced symmetry. AVHE requires breaking the enforced symmetry to break valley degeneracy or induce nonzero net Berry curvature. For example, for the DMSG $P321$, a zero net Berry curvature will be enforced by the twofold rotational symmetry $2_{100}$, which need to be broken to realize AVHE. Hence, we obtain a symmetry rule for realizing the AVHE, i.e., all the enforced symmetry operators connecting the two valleys or reversing the sign of Berry curvature must be broken. According to this rule, five DMSGs ($P3$, $P312'$, $P3m'1$, $P\bar{6}$, $P\bar{6}m'2'$) permitting AVHE are identified and denoted as $G_v$. These five DMSGs indicates that AVHE can be realized in either ferromagnetic (FM) or antiferromagnetic (AFM) systems. Meanwhile, the other 62 DMSGs of 2D hexagonal lattices do not permit AVHE and are denoted as $G_{nv}$ (see Table S1 [34]).

Although AVHE is prohibited in a system with $G_{nv}$, it can be induced by transforming the group $G_{nv}$ to $G_v$ via breaking enforced symmetries, including applying an external electric field or bilayer stacking. We derive nine groups ($P\bar{3}$, $P\bar{3}'$, $P312$, $P321$, $P32'1$, $P\bar{3}'m'1$, $P\bar{3}m'1$, $P\bar{6}'$, $P\bar{6}'m'2$) allowing for the realization of AVHE under an out-of-plane electric field (Sec. 2 of the Supplemental Material [34]). In the case of the bilayer stacking, we consider the homobilayer (B) with the corresponding single layer (S) and $B = S + \hat{O}S$, where $\hat{O} = \{O|t_o\}$ is a stacking operator with rotational part $O$ and translational part $t_o$ [35]. We focus on the single-layer ferromagnet with out-of-plane magnetic moment, and the stacking bilayer with interlayer FM or AFM orderings. We consider a high symmetric stacking bilayer with the interlayer translation $t_o = (\frac{1}{3}, \frac{2}{3})$ or $(\frac{2}{3}, \frac{1}{3})$, where the trivial out-of-plane translation is omitted. Its DMSG belongs to one of the groups $P3$, $P\bar{3}$, $P32'1$, $P3m'1$, $P\bar{3}m'1$ for the interlayer FM ordering or one of the groups $P3$, $P\bar{3}'$, $P321$, $P3m'1$, $P\bar{3}'m'1$ for the interlayer AFM ordering, indicating that it can exhibit a spontaneous AVHE or an induced AVHE under an electric field. Hence, AVHE can be induced in a ferromagnetic monolayer either by stacking or by a combination of stacking and an electric field perpendicular to the interface. Besides, AVHE can be induced in heterobilayers only when the group of heterobilayer belongs to $P3$ or $P3m'1$ (Sec. 2 of the Supplemental Material [34]).

Having identified the DMSGs allowing the realization of AVHE, we then explore how to manipulate the sign of Hall voltage in the AVHE. Reversing the sign of Hall voltage requires reversing the sign of Berry curvature, i.e., switching the valley DOF. We consider the switching between two valley-polarized states, state I and state II (lower panel of Fig. 1), of a system which are connected by an asymmetry operator, referred to as the connection operator ($\hat{N}_m$, the subscript "m" denotes "magnetic"). In other words, state II is a transformation of state I under $\hat{N}_m$ operation. We denote $\hat{N}_m$ as $\hat{N}_m = \{N_m|t_{N_m}\}$, with the rotational part $N_m \in O(3) \otimes \{1, 1'\}$ and the translational part $t_{N_m}$. Under the asymmetry operation $\hat{N}_m$, each physical quantity in Table I follows the same transformation rule as that under the symmetry operation $\hat{R}_m$ (Sec. 3 of the Supplemental Material [34]). We obtain the rule for switching the valley DOF, i.e., a connection operation $\hat{N}_m$ is required to reverse the sign of Berry curvature. As the spin and net magnetization transform in the same way as Berry curvature, one can switch valley DOF via reversing spin and net magnetization through magnetic field. Particularly, when $\widehat{N}_m$ can simultaneously reverse Berry curvature and layer (electric) polarization, valley DOF can also be switched by reversing layer (electric) polarization via electric field. Obviously, the connection operation $\widehat{N}_m$ determines the manipulation methods of valley DOF, the coupling between valley, spin and layer DOFs, as well as the magnetoelectric coupling.

Theoretically, there are infinite $\widehat{N}_m$ operations that can reverse Berry curvature and thus switch valley DOF. For example, one can switch valley DOF by combining an arbitrary rotation along $z$ axis and a time-reversal operation. However, only some of $\widehat{N}_m$ operations are easy to be realized by reversing the magnetization, reversing the electric field or interlayer sliding. Reversing the out-of-plane magnetic moment can switch valley DOF since $N_m = 1'$ reverses the sign of Berry curvature. Meanwhile, flipping the out-of-plane electric field can switch valley DOF only when the corresponding $\widehat{N}_m$ reverses the sign of Berry curvature (Sec. 3 of the Supplemental Material [34]). Additionally, for the switching of stacking orders, we consider two homobilayers with interlayer translation $\boldsymbol{t}_o = (\frac{1}{3}, \frac{2}{3})$ and $(\frac{2}{3}, \frac{1}{3})$, and assume that the direction of magnetic moment remains unchanged in the process of interlayer sliding. We establish an equation set for $\widehat{N}_m$ that connects two stacking bilayers. By solving the equation set for each layer group of composing monolayer, we identify all the stacking bilayers connected by $\widehat{N}_m$ (Table S4 of the Supplemental Material [34]). When the two bilayers have interlayer FM ordering, the valley DOF cannot be switched by sliding because their connection operation $\widehat{N}_m$ is forced to preserve the out-of-plane magnetization and thus preserve the sign of Berry curvature. Only when they have interlayer AFM ordering and spontaneous AVHE, as well as $N_m \in \{2_{120}, m'_{001}\}G_3$, valley DOF can be switched by changing the stacking order. Besides, the valley DOF in heterobilayers can be switched by reversing the magnetization or changing the stacking order (Sec. 3 of the Supplemental Material [34]).

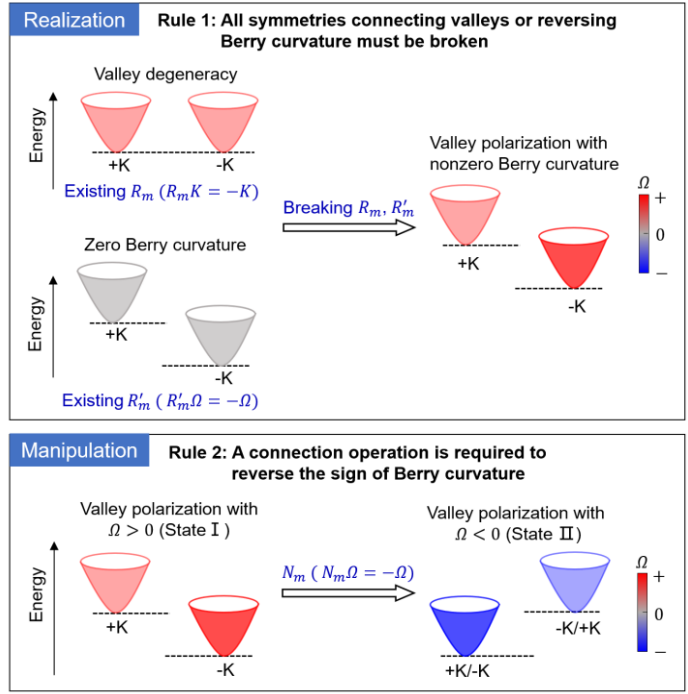

FIG. 1. The upper panel schematically illustrates the rule (Rule 1) for the realization of AVHE. The left part are two examples prohibiting AVHE, while the right one permits AVHE. The AVHE can be either spontaneous or induced by applying an electric field or stacking. K is the high symmetry point in reciprocal space with the fractional coordinate $(1/3, 1/3)$. $R_m$ is the rotational part (point group symmetry) of $\hat{R}_m$. The red, grey and blue colors denote the positive, zero and negative net Berry curvatures, respectively. The lower panel schematically illustrates the rule (Rule 2) for the manipulation of AVHE. The two valley-polarized states with positive (state I) and negative (state II) Berry curvatures are connected by an asymmetry operator $\hat{N}_m$. $N_m$ is the rotational part of $\hat{N}_m$. The manipulation methods include reversing the magnetization, reversing the electric field and interlayer sliding. K point maintains or reverses sign under $\hat{N}_m$ operation.

TABLE I. Transformations under symmetry operations in 2D hexagonal lattices. The notations of operations follow the convention on the Bilbao Crystallographic Server [33]. The specific elements of the operation set $R_m G_3$ are listed in Table S2 [34], where $G_3 = \{1, 3_{001}^+, 3_{001}^-, {}^d1, {}^d3_{001}^+, {}^d3_{001}^-\}$ denotes the double magnetic point group 3. $s_z$, $\Omega$ and $P$ are the simplified denotations for $z$ components of the spin vector, Berry curvature and electric polarization, respectively. $L$ is the layer polarization [36]. "+"("−") denotes the sign preservation (reversal) of a physical quantity under an operation. "✓" ("✗") denotes that the AVHE is symmetry allowed (forbidden).

| $R_m, N_m$ | $K$ | $s_z, \Omega$ | $L, P$ | AVHE | $R_m, N_m$ | $K$ | $s_z, \Omega$ | $L, P$ | AVHE |
|---|---|---|---|---|---|---|---|---|---|
| $1G_3$ | + | + | + | ✓ | $1'G_3$ | − | − | + | ✗ |
| $2_{001}G_3$ | − | + | + | ✗ | $2'_{001}G_3$ | + | − | + | ✗ |
| $2_{100}G_3$ | + | − | − | ✗ | $2'_{100}G_3$ | − | + | − | ✗ |
| $2_{120}G_3$ | − | − | − | ✗ | $2'_{120}G_3$ | + | + | − | ✓ |
| $\bar{1}G_3$ | − | + | − | ✗ | $\bar{1}'G_3$ | + | − | − | ✗ |
| $m_{001}G_3$ | + | + | − | ✓ | $m'_{001}G_3$ | − | − | − | ✗ |
| $m_{100}G_3$ | − | − | + | ✗ | $m'_{100}G_3$ | + | + | + | ✓ |
| $m_{120}G_3$ | + | − | + | ✗ | $m'_{120}G_3$ | − | + | + | ✗ |

In the above discussions, we focus on the case that the valley is located at $K$ or $-K$ point. Now we generalize the AVHE rules to generic $\boldsymbol{k}$ points. Under the symmetries of $G_3$ in 2D hexagonal lattices, the energy level and Berry curvature remain unchanged. The valleys protected by above symmetries are defined as equivalent valleys. Then the rule for realizing AVHE can be generalized to that all the enforced symmetries connecting nonequivalent valleys or reversing the sign of Berry curvature must be broken. And the rule for manipulating AVHE remains unchanged. Based on these generalized rules, we can derive the same five groups ($P3$, $P312'$, $P3m'1$, $P\bar{6}$, $P\bar{6}m'2'$) for the high symmetry line $\Gamma K$ or $KM$ as that of $K$ point. Hence, the results of transforming $G_{nv}$ to $G_v$ and switching valley DOF for $K$ point can be directly applied to the case of $\Gamma K$ or $KM$. Besides, the AVHE can be realized in groups $P3$, $P32'1$, $P31m'$, $P\bar{6}$, $P\bar{6}2'm'$ along the high symmetry line $\Gamma M$, and be realized in group $P3$ for other generic $\boldsymbol{k}$ points. The valley DOF in these cases can also be switched by reversing the magnetization, reversing the electric field or interlayer sliding (Sec. 4 of the Supplemental Material [34]).

*Spontaneous AVHE in AgCrP$_2$Se$_6$ and CrOBr.*—We apply the above AVHE rules to the well-known AVHE systems with space group (SG) $P\bar{6}m2$. For example, the ferromagnetic monolayer 2H-VSe$_2$ with SG $P\bar{6}m2$ and an out-of-plane magnetic moment [8] possesses the DMSG $G_v = P\bar{6}m'2'$, and thus exhibit the AVHE. The valley DOF can be switched by reversing the magnetic moment as the connection operation $N_m = 1'$ reverses the sign of Berry curvature. For bilayer 2H-VS$_2$ [27], we reveal that $N_m = 2_{120}$ determines the switching of valley DOF and the coupling of ferrovalley and ferroelectricity (Sec. 5 of the Supplemental Material [34]). We then apply AVHE rules to perform a high-throughput screening for AVHE materials. Ten new spontaneous AVHE candidates (i.e., ferrovalley candidates) are identified from Ref. [37] (Sec. 6 of the Supplemental Material [34]). Among them, the monolayer AgCrP$_2$Se$_6$ with SG $P312$ and an out-of-plane magnetic moment possesses the DMSG $G_v = P312'$. Density functional theory (DFT) calculations verify that valley polarization with a nonzero Berry curvature appears in the conduction band when the magnetization of monolayer AgCrP$_2$Se$_6$ is upward, confirming the realization of AVHE [Fig. 2(a)]. By reversing the magnetic moment, the valley DOF is switched because the sign of Berry curvature is reversed by $N_m = 1'$ [Fig. 2(b)]. We also verify that the AVHE exists in monolayer CrOBr whose valleys of the valence band are located along $\Gamma K$ line (see details in Sec. 6 of the Supplemental Material [34]).

*Induced AVHE in FeCl$_2$, TcGeSe$_3$.*—We also apply the AVHE rules to magnetic systems with SG $P\bar{3}m1$ and $P\bar{3}1m$, and verify the rules using DFT calculations. Monolayer FeCl$_2$ with SG $P\bar{3}m1$ has been experimentally synthesized [38,39]. It has out-of-plane magnetic moment [40,41] and thus has the DMSG $P\bar{3}m'1 = G_v \otimes \{1, \bar{1}\}$, where $G_v = P3m'1$. According to the rules shown in Fig. 1, it does not have AVHE, but AVHE can be induced by applying an electric field and reversed by switching magnetic moment. The valleys in $K$ and $K' = -K$ are energetically degenerate due to symmetry $\bar{1}G_v$. Besides, $\bar{1}G_v$ also gives rise to opposite layer polarizations at $K$ and $-K$ valleys, where the layer polarization refers to the polarization between the top and bottom Cl atomic layers according to the definition in Ref. [36] (see details in the Supplemental Material [34]). Under an out-of-plane electric field [Fig. 2(c)], the symmetry $\bar{1}G_v$ is broken. The group $G_{nv} = P\bar{3}m'1$ transforms into $G_v = P3m'1$, enabling the realization of AVHE. Since $-K$ valley has a positive layer polarization, its energy level will gain a lower electrostatic potential and move downward compared to that of $K$ valley under the downward electric field [7,36] [the middle panel of Fig. 2(c)]. The valley polarization results in a slight polarization of the Berry curvature between $K$ and $-K$ valleys (Fig. S6 of the Supplemental Material [34]). Having achieved the AVHE, we then consider how to switch it. The valley DOF cannot be switched by flipping the direction of electric field as the connection operation $N_m = \bar{1}$ preserves the sign of Berry curvature. Hence, we must flip the direction of magnetic moment [Fig. 2(d)], since $N_m = 1'$ guarantees the sign reversal of Berry curvature and anomalous Hall conductivity without changing their amplitudes [the lower panel of Fig. 2(d), Fig. S6 and Fig. S7]. The amplitude of valley splitting (12 meV) also remains unchanged due to $N_m = 1'$.

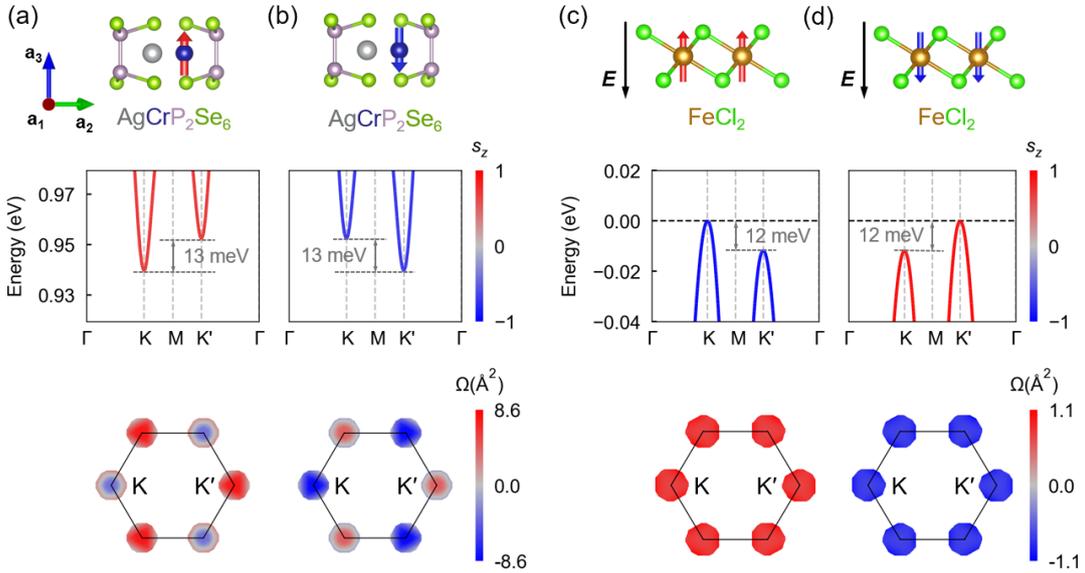

FIG. 2. The configurations, band structures and Berry curvatures for monolayer AgCrP$_2$Se$_6$ with (a) an upward and (b) downward magnetic moment, for monolayer FeCl$_2$ (c) with an upward and (d) downward magnetic moment under an external electric field ($E$) of $-0.05$ V/Å . In all upper panels, the red (blue) arrow denotes the upward (downward) magnetic moment. The black arrow denotes the electric field. In all middle panels, $K'$ denotes $-K$, and the Fermi level is set to zero. The red (blue) lines denote the bands contributed by upward (downward) spin. Lower panels show the Berry curvature distributions around $K$ and $-K$ valleys for the valence band. The red (blue) color denotes the positive (negative) Berry curvature.

Monolayer TcGeSe$_3$ with SG $P\bar{3}1m$ also has out-of-plane magnetic moment [42],

described by the group $G_{nv} = P\bar{3}1m'$. According to the rules in Fig. 1, AVHE is prohibited and cannot be induced by an electric field, in contrast to monolayer $FeCl_2$. But AVHE can be realized by a combination of bilayer stacking and an external electric field, and can be further reversed via flipping the electric field when the bilayer has an interlayer AFM ordering. We stack $TcGeSe_3$ bilayers with the stacking operator $\hat{O} = \{m_{001}|t\}$ and perform DFT calculations to identify the magnetic ground state and AVHE. The bilayer with $t = (\frac{1}{3}, \frac{2}{3})$ has interlayer AFM ground state [Fig. 3(a)] and thus has the DMSG $P321 = G_v \cup 2_{100}G_v$, where $G_v = P3$. AVHE is prohibited because the conduction band is doubly degenerated at $K$ point [the middle panel of Fig. 3(a)], with zero net Berry curvature enforced by symmetries $2_{100}G_v$ [the lower panel of Fig. 3(a)]. Applying an out-of-plane electric field breaks the symmetry $2_{100}G_v$ and transforms the group $G_{nv} = P321$ into $G_v = P3$ [Fig. 3(b)]. The electric field lifts the band degeneracy and induces a AVHE [the middle and lower panels of Fig. 3(b)]. The Berry curvature also exhibits polarization between $K$ and $-K$ valleys (Fig. S9 of the Supplemental Material [34]). By flipping the electric field, valley DOF is switched since $N_m = 2_{100}$ reverses the signs of Berry curvature and anomalous Hall conductivity without changing their amplitude [the middle and lower panels of Fig. 3(c), Fig. S9 and Fig. S10]. The amplitude of valley splitting (11 meV) also remains unchanged due to $N_m = 2_{100}$. Meanwhile, the spin and layer DOFs are also switched, indicating that the switchable anomalous Hall effect is locked with valley, spin and layer DOFs [43,44]. However, valley DOF cannot be switched by shifting top layer with respect to bottom layer with $t = (\frac{1}{3}, -\frac{1}{3})$ as $N_m = m'_{120}$ preserves the sign of the out-of-plane magnetization and thus the sign of Berry curvature under the electric field. Besides, for another stacking $TcGeSe_3$ bilayer with $\hat{O} = \{1|(\frac{1}{3}, \frac{2}{3})\}$, it has the interlayer AFM ground state with $G_{nv} = P\bar{3}$. Its conduction band is doubly degenerated at all $\boldsymbol{k}$ points and has a zero net Berry curvature and thus a vanishing AVHE due to the symmetry $\bar{1}'$. AVHE can also be induced by applying an out-of-plane electric field and reversed by flipping the field (see details in Sec. 6 of the Supplemental Material [34]).

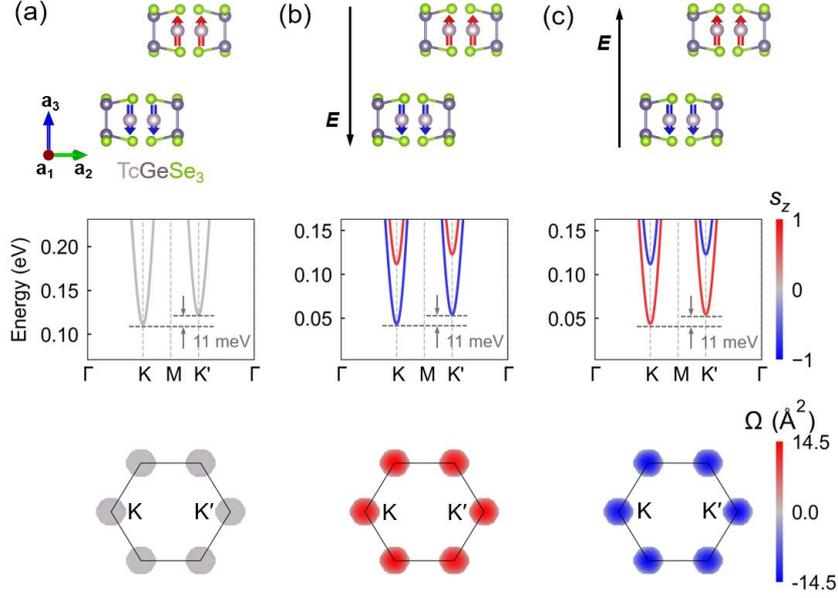

FIG. 3. The configurations, band structures and Berry curvatures of AFM bilayer TcGeSe$_3$ stacked by operator $\hat{O} = \{m_{001}|(\frac{1}{3},\frac{2}{3})\}$ (a) without an external electric field, (b) with a downward ($E = -0.01$ V/Å) and (c) upward external electric field ($E = 0.01$ V/Å). In all upper panels, the red (blue) arrow denotes the upward (downward) magnetic moment. The black arrow denotes the electric field. In all middle panels, $K'$ denotes $-K$, and the Fermi level is set to zero. The red (blue) lines denote the bands contributed by upward (downward) spin. The grey color denotes zero spin polarization. Lower panels show the Berry curvature distributions around $K$ and $-K$ valleys for the conduction band. The red, grey and blue colors denote the positive, zero and negative net Berry curvature, respectively.

*Summary and discussion.*—In conclusion, we provide general rules for realizing AVHE and switching valley DOF. Based on symmetry analysis, we identify all DMSGs permitting AVHE and reveal that the connection operation determines the switching of valley DOF. In fact, the AVHE rules are also applicable to hexagonal lattices where threefold rotational symmetry is broken by magnetic order, such as in-plane magnetized ferromagnets, which are not previously expected to exhibit the AVHE [18,45] (Sec. 7 of the Supplemental Material [34]). AVHE rules are also applicable to other 2D and 3D lattices (Sec. 7 [34]). Besides, AVHE materials can be utilized to construct magnetic tunnel junctions, enabling enhanced tunnel magnetoresistance due to valley-spin coupling (Sec. 8 of the Supplemental Material [34]). Moreover, beyond the valley DOF, the connection operation $\widehat{N}_m$ also dictates the switching of the spin DOF, layer DOF, electric polarization, and magnetization. Consequently, $\widehat{N}_m$ determines the interplay between microscopic electronic DOFs and macroscopic electric and magnetic properties. We can employ $\widehat{N}_m$ to design systems with switchable or coupled physical properties in spintronics, layertronics, magnetoelectrics, multiferroics, etc [46-48]. Our findings not only provide a unified framework for designing AVHE systems, but also offers the connection operation for designing systems with switchable or coupled physical properties in other fields.

The authors thank Bo-Xuan Li for helpful discussions. This work is supported by funds from the National Key Research and Development Program of China (2022YFA1204100, 2021YFA1201501), the National Natural Science Foundation of China (62488201, 52272172, 12374172) and the Major Program of the National Natural Science Foundation of China (92163206).

*Corresponding authors: jtsun@bit.edu.cn, sxdu@iphy.ac.cn

[1] J. R. Schaibley, H. Y. Yu, G. Clark, P. Rivera, J. S. Ross, K. L. Seyler, W. Yao, and X. D. Xu, Nat. Rev. Mater. **1**, 16055 (2016).
[2] X.-W. Shen, H. Hu, and C.-G. Duan, in *Spintronic 2D Materials* (Elsevier, 2020), pp. 65.
[3] A. Rycerz, J. Tworzydło, and C. W. J. Beenakker, Nat. Phys. **3**, 172 (2007).
[4] D. Xiao, W. Yao, and Q. Niu, Phys. Rev. Lett. **99**, 236809 (2007).
[5] D. Xiao, G. B. Liu, W. Feng, X. Xu, and W. Yao, Phys. Rev. Lett. **108**, 196802 (2012).
[6] K. F. Mak, D. Xiao, and J. Shan, Nat. Photonics **12**, 451 (2018).
[7] W.-Y. Tong and C.-G. Duan, npj Quantum Mater. **2**, 47 (2017).
[8] W. Y. Tong, S. J. Gong, X. Wan, and C. G. Duan, Nat. Commun. **7**, 13612 (2016).
[9] S.-D. Guo, W. Xu, Y. Xue, G. Zhu, and Y. S. Ang, Phys. Rev. B **109**, 134426 (2024).
[10] S.-D. Guo, Y.-L. Tao, Z.-Y. Zhuo, G. Zhu, and Y. S. Ang, Phys. Rev. B **109**, 134402 (2024).
[11] R.-C. Zhang and H.-Y. Chen, Phys. Rev. B **108**, 075423 (2023).
[12] S. Ji, R. Yao, C. Quan, J. Yang, F. Caruso, and X. a. Li, Phys. Rev. B **107**, 174434 (2023).
[13] Y. Zhu, Q. Cui, Y. Ga, J. Liang, and H. Yang, Phys. Rev. B **105**, 134418 (2022).
[14] P. Zhao, Y. Dai, H. Wang, B. Huang, and Y. Ma, ChemPhysMater **1**, 56 (2022).

[15] R. J. Sun, R. Liu, J. J. Lu, X. W. Zhao, G. C. Hu, X. B. Yuan, and J. F. Ren, Phys. Rev. B **105**, 235416 (2022).

[16] W. H. Du, R. Peng, Z. L. He, Y. Dai, B. B. A. Huang, and Y. D. Ma, npj 2D Mater. Appl. **6**, 11 (2022).

[17] S.-D. Guo, J.-X. Zhu, W.-Q. Mu, and B.-G. Liu, Phys. Rev. B **104**, 224428 (2021).

[18] R. Peng, Y. Ma, X. Xu, Z. He, B. Huang, and Y. Dai, Phys. Rev. B **102**, 035412 (2020).

[19] T. Zhang, X. Xu, B. Huang, Y. Dai, and Y. Ma, npj Comput. Mater. **8**, 64 (2022).

[20] D. Xiao, M.-C. Chang, and Q. Niu, Rev. Mod. Phys. **82**, 1959 (2010).

[21] N. Nagaosa, J. Sinova, S. Onoda, A. H. MacDonald, and N. P. Ong, Rev. Mod. Phys. **82**, 1539 (2010).

[22] L. Du, T. Hasan, A. Castellanos-Gomez, G.-B. Liu, Y. Yao, C. N. Lau, and Z. Sun, Nat. Rev. Phys. **3**, 193 (2021).

[23] Z. Gong, G. B. Liu, H. Yu, D. Xiao, X. Cui, X. Xu, and W. Yao, Nat. Commun. **4**, 2053 (2013).

[24] A. Zhang, K. Yang, Y. Zhang, A. Pan, and M. Chen, Phys. Rev. B **104**, 201403 (2021).

[25] P. Zhao, Y. Ma, C. Lei, H. Wang, B. Huang, and Y. Dai, Appl. Phys. Lett. **115**, 261605 (2019).

[26] L. Feng, X. Chen, and J. Qi, Phys. Rev. B **108**, 115407 (2023).

[27] X. Liu, A. P. Pyatakov, and W. Ren, Phys. Rev. Lett. **125**, 247601 (2020).

[28] J. Ma, X. Luo, and Y. Zheng, npj Comput. Mater. **10**, 102 (2024).

[29] Y. Z. Wu, J. W. Tong, L. Deng, F. F. Luo, F. B. Tian, G. W. Qin, and X. M. Zhang, Nano Lett. **23**, 6226 (2023).

[30] Y. Q. Li, X. Zhang, X. Shang, Q. W. He, D. S. Tang, X. C. Wang, and C. G. Duan, Nano Lett. **23**, 10013 (2023).

[31] L. Elcoro, B. J. Wieder, Z. Song, Y. Xu, B. Bradlyn, and B. A. Bernevig, Nat. Commun. **12**, 5965 (2021).

[32] G.-B. Liu, Z. Zhang, Z.-M. Yu, and Y. Yao, Comput. Phys. Commun. **288**, 108722 (2023).

[33] L. Elcoro *et al.*, J. Appl. Crystallogr. **50**, 1457 (2017).

[34] See Supplemental Material for details about group theory analysis, AVHE in known systems, computational methods, high-throughput screening for AVHE materials, DFT results of monolayer $AgCrP_2Se_6$, CrOBr, $FeCl_2$ and bilayer $TcGeSe_3$, AVHE in other systems, and AVHE materials for enhancing tunnel magnetoresistance.

[35] J. Ji, G. Yu, C. Xu, and H. Xiang, Phys. Rev. Lett. **130**, 146801 (2023).

[36] Z. M. Yu, S. Guan, X. L. Sheng, W. Gao, and S. A. Yang, Phys. Rev. Lett. **124**, 037701 (2020).

[37] J. Sødequist and T. Olsen, npj Comput. Mater. **10**, 170 (2024).

[38] X. Zhou *et al.*, J. Phys. Chem. C **124**, 9416 (2020).

[39] S. Jiang *et al.*, ACS Nano **17**, 363 (2022).

[40] A. S. Botana and M. R. Norman, Phys. Rev. Mater. **3**, 044001 (2019).

[41] Q. Yao, J. Li, and Q. Liu, Phys. Rev. B **104**, 035108 (2021).

[42] J.-Y. You, Z. Zhang, X.-J. Dong, B. Gu, and G. Su, Phys. Rev. Res. **2**, 013002

(2020).

[43] A. Gao *et al.*, Nature **595**, 521 (2021).

[44] T. Zhang, X. Xu, B. Huang, Y. Dai, L. Kou, and Y. Ma, Mater. Horiz. **10**, 483 (2023).

[45] Q. Cui, Y. Zhu, J. Liang, P. Cui, and H. Yang, Phys. Rev. B **103**, 085421 (2021).

[46] L. Šmejkal, R. González-Hernández, T. Jungwirth, and J. Sinova, Sci. Adv. **6**, eaaz8809 (2020).

[47] D.-F. Shao, J. Ding, G. Gurung, S.-H. Zhang, and E. Y. Tsymbal, Phys. Rev. Appl. **15** (2021).

[48] T. Cao, D.-F. Shao, K. Huang, G. Gurung, and E. Y. Tsymbal, Nano Lett. **23**, 3781 (2023).